\documentstyle[twoside]{article}
\newcommand{\avg}[1]{\langle {#1} \rangle}

\catcode`\@=11
\long\def\@makefntext#1{
\protect\noindent \hbox to 3.2pt {\hskip-.9pt  
$^{{\eightrm\@thefnmark}}$\hfil}#1\hfill}		

\def\thefootnote{\fnsymbol{footnote}}
\def\@makefnmark{\hbox to 0pt{$^{\@thefnmark}$\hss}}	
	
\def\ps@myheadings{\let\@mkboth\@gobbletwo
\def\@oddhead{\hbox{}
\rightmark\hfil\eightrm\thepage}   
\def\@oddfoot{}\def\@evenhead{\eightrm\thepage\hfil
\leftmark\hbox{}}\def\@evenfoot{}
\def\sectionmark##1{}\def\subsectionmark##1{}}

\oddsidemargin=\evensidemargin
\renewcommand{\thefootnote}{\fnsymbol{footnote}}

\newcounter{sectionc}\newcounter{subsectionc}\newcounter{subsubsectionc}
\renewcommand{\section}[1] {\vspace{12pt}\addtocounter{sectionc}{1} 
\setcounter{subsectionc}{0}\setcounter{subsubsectionc}{0}\noindent 
	{\tenbf\thesectionc. #1}\par\vspace{5pt}}
\renewcommand{\subsection}[1] {\vspace{12pt}\addtocounter{subsectionc}{1} 
	\setcounter{subsubsectionc}{0}\noindent 
	{\bf\thesectionc.\thesubsectionc. {\kern1pt \bfit #1}}\par\vspace{5pt}}
\renewcommand{\subsubsection}[1] {\vspace{12pt}\addtocounter{subsubsectionc}{1}
	\noindent{\tenrm\thesectionc.\thesubsectionc.\thesubsubsectionc.
	{\kern1pt \tenit #1}}\par\vspace{5pt}}
\newcommand{\nonumsection}[1] {\vspace{12pt}\noindent{\tenbf #1}
	\par\vspace{5pt}}

\newcounter{appendixc}
\newcounter{subappendixc}[appendixc]
\newcounter{subsubappendixc}[subappendixc]
\renewcommand{\thesubappendixc}{\Alph{appendixc}.\arabic{subappendixc}}
\renewcommand{\thesubsubappendixc}
	{\Alph{appendixc}.\arabic{subappendixc}.\arabic{subsubappendixc}}

\renewcommand{\appendix}[1] {\vspace{12pt}
        \refstepcounter{appendixc}
        \setcounter{figure}{0}
        \setcounter{table}{0}
        \setcounter{lemma}{0}
        \setcounter{theorem}{0}
        \setcounter{corollary}{0}
        \setcounter{definition}{0}
        \setcounter{equation}{0}
        \renewcommand{\thefigure}{\Alph{appendixc}.\arabic{figure}}
        \renewcommand{\thetable}{\Alph{appendixc}.\arabic{table}}
        \renewcommand{\theappendixc}{\Alph{appendixc}}
        \renewcommand{\thelemma}{\Alph{appendixc}.\arabic{lemma}}
        \renewcommand{\thetheorem}{\Alph{appendixc}.\arabic{theorem}}
        \renewcommand{\thedefinition}{\Alph{appendixc}.\arabic{definition}}
        \renewcommand{\thecorollary}{\Alph{appendixc}.\arabic{corollary}}
        \renewcommand{\theequation}{\Alph{appendixc}.\arabic{equation}}
        \noindent{\tenbf Appendix#1}\par\vspace{5pt}}
\newcommand{\subappendix}[1] {\vspace{12pt}
	\newcommand{\avg}[1]{\langle {#1} \rangle}
        \refstepcounter{subappendixc}
        \noindent{\bf Appendix \thesubappendixc. {\kern1pt \bfit #1}}
	\par\vspace{5pt}}
\newcommand{\subsubappendix}[1] {\vspace{12pt}
        \refstepcounter{subsubappendixc}
        \noindent{\rm Appendix \thesubsubappendixc. {\kern1pt \tenit #1}}
	\par\vspace{5pt}}

\newcommand{\textlineskip}{\baselineskip=13pt}
\newcommand{\smalllineskip}{\baselineskip=10pt}

\def\eightcirc{
\begin{picture}(0,0)
\put(4.4,1.8){\circle{6.5}}
\end{picture}}
\def\eightcopyright{\eightcirc\kern2.7pt\hbox{\eightrm c}} 

\newcommand{\copyrightheading}[1]
	{\vspace*{-2.5cm}\smalllineskip{\flushleft
	{\footnotesize Mathematical Models and Methods in Applied Sciences #1}\\
	{\footnotesize $\eightcopyright$\, World Scientific Publishing
	 Company}\\
	 }}

\def\abstracts#1#2#3{{
	\centering{\begin{minipage}{4.5in}\baselineskip=10pt\footnotesize
	\parindent=0pt #1\par 
	\parindent=15pt #2\par
	\parindent=15pt #3
	\end{minipage}}\par}} 

\newcounter{itemlistc}
\newcounter{romanlistc}
\newcounter{alphlistc}
\newcounter{arabiclistc}

\newcommand{\fcaption}[1]{
        \refstepcounter{figure}
        \setbox\@tempboxa = \hbox{\footnotesize Fig.~\thefigure. #1}
        \ifdim \wd\@tempboxa > 5in
           {\begin{center}
        \parbox{5in}{\footnotesize\smalllineskip Fig.~\thefigure. #1}
            \end{center}}
        \else
             {\begin{center}
             {\footnotesize Fig.~\thefigure. #1}
              \end{center}}
        \fi}

\newcommand{\tcaption}[1]{
        \refstepcounter{table}
        \setbox\@tempboxa = \hbox{\footnotesize Table~\thetable. #1}
        \ifdim \wd\@tempboxa > 5in
           {\begin{center}
        \parbox{5in}{\footnotesize\smalllineskip Table~\thetable. #1}
            \end{center}}
        \else
             {\begin{center}
             {\footnotesize Table~\thetable. #1}
              \end{center}}
        \fi}

\def\@citex[#1]#2{\if@filesw\immediate\write\@auxout
	{\string\citation{#2}}\fi
\def\@citea{}\@cite{\@for\@citeb:=#2\do
	{\@citea\def\@citea{,}\@ifundefined
	{b@\@citeb}{{\bf ?}\@warning
	{Citation `\@citeb' on page \thepage \space undefined}}
	{\csname b@\@citeb\endcsname}}}{#1}}

\newif\if@cghi
\def\cite{\@cghitrue\@ifnextchar [{\@tempswatrue
	\@citex}{\@tempswafalse\@citex[]}}
\def\citelow{\@cghifalse\@ifnextchar [{\@tempswatrue
	\@citex}{\@tempswafalse\@citex[]}}
\def\@cite#1#2{{$\null^{#1}$\if@tempswa\typeout
	{IJCGA warning: optional citation argument 
	ignored: `#2'} \fi}}

\def\pmb#1{\setbox0=\hbox{#1}
	\kern-.025em\copy0\kern-\wd0
	\kern.05em\copy0\kern-\wd0
	\kern-.025em\raise.0433em\box0}

\def\fnt#1#2{\footnotetext{\kern-.3em
	{$^{\mbox{\scriptsize #1}}$}{#2}}}

\def\fpage#1{\begingroup
\voffset=.3in
\thispagestyle{empty}\begin{table}[b]\centerline{\footnotesize #1}
	\end{table}\endgroup}

\font\tenrm=cmr10
\font\tenit=cmti10 
\font\tenbf=cmbx10
\font\bfit=cmbxti10 at 10pt
\font\ninerm=cmr9

\font\eightrm=cmr8

\def\qed{\hbox{${\vcenter{\vbox{			
   \hrule height 0.4pt\hbox{\vrule width 0.4pt height 6pt
   \kern5pt\vrule width 0.4pt}\hrule height 0.4pt}}}$}}

\renewcommand{\thefootnote}{\fnsymbol{footnote}}	

\def\theequation{\thesectionc.\arabic{equation}}	

\begin{document}

\normalsize\textlineskip
\thispagestyle{empty}
\setcounter{page}{1}

\copyrightheading{}			

\vspace*{0.88truein}

\fpage{1}
\centerline{\bf DRIVING FORCE IN INVESTMENT}
\vspace*{0.37truein}
\centerline{\footnotesize ANDREA CAPOCCI and YI-CHENG ZHANG}
\vspace*{0.015truein}
\centerline{\footnotesize\it Institut de Physique Th\'{e}orique, Universit\'{e} de Fribourg, CH-1700, Switzerland}
\baselineskip=10pt
\centerline{\footnotesize\it E-mail: Andrea.Capocci@unifr.ch}
\centerline{\footnotesize\it E-mail: Yi-Cheng.Zhang@unifr.ch}
\vspace*{10pt}
\vspace*{0.225truein}
\vspace*{0.21truein}
\abstracts{We study investment strategy in different models of financial markets, where the investors cannot reach a perfect knowledge about available assets. The investor spends a certain effort to get information; this allows him to better choose the investment strategy, and puts a selective pressure upon assets. The best strategy is then a compromise between diversification and effort to get information.
}{}{}



\vspace*{1pt}\textlineskip	
\section{Introduction}  	
\vspace*{-0.5pt}
\noindent
Standard theory in finance assumes perfect information. For example, portfolio theories and CAPM (Capital Asset Pricing Model) reduce investment strategies to a mechanic optimization problem, with a given set of parameters like expected returns and volatility (see [2]). In real life the property of an asset, say a stock, cannot be characterized simply by a few parameters. In fact the possible sources of influence on a stocks price must be unlimited. On the other hand, not all parameters are as important and nobody has infinite effort to study all of them.

We present an alternative study of investment strategies. We consider all the information is imperfect, but upon the investor's diligent effort, the precision can be improved. But there is a limit: even with infinite effort precise prediction is not possible. This is similar in spirit with the approach followed by Grossman and Stiglitz [4], but our approach is simpler and has wider applicability. The important difference separates ours from [4] is that we do not equalize information cost and its monetary value. We use instead the concept of {\it effort}, which cannot always be bought---having to do with an investor's experience and quality. These human factors were consistently omitted in economics literature.

We use two sets of parameters (i.e. information): one is the reality which can never be known exactly; the other the perceived values which are always distorted by investors. With increased effort an investor can improve the precision of information, leading to better gains. Not only an investor gains, thanks to his diligent effort, he does also a social service by putting pressure on the worse performing opportunity (e.g. a stock), whereas encouraging the better opportunity to grow with increased capital. In a sense every increase in effort puts additional evolutionary pressure among the opportunities. 

Human effort is always finite. Be it a big investment bank or a day trader, the effort (including monetary cost) that can be used to search out more precise information about investment opportunities. If this sounds evident, then we are led to the inevitable conclusion that a rational investor should not diversify too much. The logic runs as follows: an investor has the total effort to spend on finding out information. There are a sufficiently large number of opportunities available. If he concentrate on only one of them, then he can expect to obtain good quality of information on that one, while remains ignorant of all the others. We know that diversification is a good thing: it reduces fluctuation with the same expected ``gain''. Thus more leverage can be used to achieve higher gain, with the same volatility. So our investor would heed the advice from standard theory to diversify---but only to a point. In fact, the rational way for him is diversify somewhat, to still obtain a reasonable estimate of the considered options. There is a compromise between the benefit of diversification and deepened knowledge.
Too much diversification dilutes his effort since he has to divide his total available effort among the alternatives. Of course the total diversification is also a strategy, if he has no effort/time to study the investment opportunities, then he is well advised to buy into an index fund.
But, on the other hand, if everybody does the same: i.e. uses no-effort strategy, then the evolutionary pressure is no longer there, and the index (i.e. the average expected gain of all the opportunities) would not even be there! 

We also consider the structure of an investment network. In investment there are so-called Favorable Games (FGs), provided by producers (see [3]). However, producers cannot keep all the potential benefit to themselves since the initial fluctuations are very large.
There is a symbiosis between the producers and speculators, in fact they form a network. The network as a whole acts as an organism of digesting the original FGs, with successive levels absorbing the initial shocks, transmitted from the producers downstream. Farther away from the initial producer level, the shocks are more and more attenuated, but the expected gains also get reduced. In a society somebody may find himself sitting comfortably at the "safer" levels, while some other might be interested in finding a place on the front "producer line". This web-like network in fact can be argued as the most efficient system in finding and exploiting the original FGs.

\textheight=7.8truein
\setcounter{footnote}{0}
\renewcommand{\thefootnote}{\alph{footnote}}

\section{Information and Diversification}
\noindent
A reasonable way to describe assets price variations (on a given time-scale) is to assume them to be multiplicative random walks with log-normal step. This comes from the assumption that growth rates of prices are more significant that their absolute variations. So, we describe the price of a financial assets as a time-dependent multiplicative random process.
We introduce a set of $N$ Gaussian random variables $x_{i}(t)$ depending on a time parameter $t$. By this set, we define $N$ independent multiplicative Gaussian random walks, whose assigned discrete time evolution is given by
\begin{equation}
p_{i}(t+1)= e^{x_{i}(t)} p_{i}(t)
\end{equation}
for $i=1,\ldots,N$, where each $x_{i}(t)$ is not correlated in time. 
To optimize an investment, one can choose different risk-return strategies. Here, by optimization we will mean the maximization of the typical capital growth rate of a portfolio. In [1] it was shown that a capital $W(t)$, invested into different financial assets who behave as multiplicative random walks, grows almost certainly at an exponential rate $\avg{\ln{\frac{W(t+1)}{W(t)}}}$, where one must average over the distribution of the single multiplicative step.
We assume that an investment is diversified according to the Kelly's optimum investment fraction, in order to maximize the typical capital growth rate over $N$ assets with identical average return $\alpha = \avg{e^{x_{i}(t)}}-1$ and squared volatility $\Delta = \avg{e^{2x_{i}(t)}} - \avg{e^{x_{i}(t)}}^2$. 
On each asset, the investor will allocate a fraction $f_{i}$ of his capital, according to the return expected from that asset. The time evolution of the total capital is ruled by the following multiplicative process
\begin{equation}
W(t+1)=[1+\sum_{i=1}^{N}f_{i}(e^{x_{i}(t)}-1)]W(t).
\end{equation}
First, we consider the case of an unlimited investment, i.e. we put no restriction to the value of $\sum_{i=1}^{N}f_{i}$. The typical growth rate
\begin{equation}
V_{typ}=\avg{\ln{[1+\sum_{i=1}^{N}f_{i}(e^{x_{i}}-1)]}}
\end{equation}
of the investor's capital can be calculated through the following 2nd-order expansion in $e^{x_{i}}-1$ if we assume that fluctuations of prices are small and uncorrelated, that seems to be quite reasonable (see also [1]):
\begin{equation} \label{vtip}
V_{typ}\simeq\sum_{i=1}^{N}[f_{i}(\avg{e^{x}}-1)-\frac{f_{i}^{2}}{2}(\avg{e^{2x}}-2\avg{e^{x}} + 1)].
\end{equation}
By solving $\frac{d}{df}V_{typ}=0$, it easy to show that the optimal value for $f_{i}$ is $f^{opt}_{i}(\alpha,\Delta)=\frac{\alpha}{\alpha^{2}+\Delta}$ for all $i$. We assume that the investor has a little ignorance about the real value of $\alpha$, that we represent by a Gaussian fluctuation around the real value of $\alpha$. In the investor's mind, each asset is different, because of this fluctuation $\alpha_{i}=\alpha+\epsilon_{i}$.
The $\epsilon_{i}$ are drawn from the same distribution, with $\avg{\epsilon_{i}}=0$ as errors are normally distributed around the real value. We suppose that the investor makes an effort $E$ to investigate and get information about the statistical parameters of the $N$ assets upon which he will spread his capital. So, his ignorance (i.e. the width of the distribution of the $\epsilon_{i}$) about the real value of $\alpha_{i}$ will be a decreasing function of the effort ``per asset'' $\frac{E}{N}$; more, we suppose that an even infinite effort will not make this ignorance vanish. In order to plug these assumptions in the model, we write the width of the distribution of $\epsilon$ as
\begin{equation}
\avg{\epsilon_{i}^{2}}=D_{0}+(\frac{N}{E})^{\gamma},
\end{equation}
with $\gamma>0$. As one can see, the greater is $E$, the more exact is the perception, and better is the investment. $D_{0}$ is the asymptotic ignorance. All the invested fraction $f^{opt}(\alpha_{i},\Delta)$ will be different, according to the investor's perception. 
Assuming that the $\epsilon_{i}$ are small, we expand all $f_{i}(\alpha+\epsilon_{i})$ in equation (\ref{vtip}) up to the 2nd order in $\epsilon_{i}$, and after averaging over the distribution of $\epsilon_{i}$, we obtain the mean value of the typical capital growth rate for an investor who provides a given effort $E$:
\begin{equation}
\overline{V_{typ}}=N[A-(D_{0}+(\frac{N}{E})^{\gamma})B]
\end{equation}
where
\begin{equation}
A = \frac{\alpha(3\Delta-\alpha^{2})}{(\Delta+\alpha^{2})^3}, B = -\frac{(\alpha^{2}-\Delta)^{2}}{2(\alpha^{2}+\Delta)^{3}}.
\end{equation}
We are now able to find the optimal number of assets to be included in the portfolio (i.e., for which the investment is more advantageous, taken into account the effort provided to get information), by solving $\frac{d}{dN}\overline{V_{typ}}=0$; it easy to see that the optimal number of assets is given by
\begin{equation}
N^{opt}(E)=E[\frac{A-D_{0}}{(1+\gamma)B}]^{\frac{1}{\gamma}},
\end{equation}
that is an increasing function of the effort $E$, as one can see in the example of Fig. 1. Notice that if the investor has no limit in the total capital fraction invested in the portfolio (so that it can be greater than 1, i.e. the investor can invest more money than he has, borrowing it from an external source), the capital can take negative values, if the assets included in the portfolio encounter a simultaneous negative step. So, if the total investment fraction is greater than 1, we should take into account also the cost of refunding loss to the bank, to predict the typical growth rate of the capital.

\section{Knowledge drives Selective Pressure}
\noindent
We suppose that the investor can choose among $N$ different option with different average returns. We assume that each asset has an average return $\alpha_{i} = \avg{e^{x_{i}}}-1$, with $i=1,\ldots,N$, and a volatility $\Delta$. As a reasonable hypothesis, we assume that in a market, good and bad stocks are equally distributed, but most of them behave in an intermediate way. We suppose that the $\alpha_{i}$ are drawn from a Gaussian distribution with mean $m$ and variance $D$, whose  density function is written $\pi(\alpha)$. Here, $m$ plays the role of a market index. Moreover, the investor's perception is not exact: given a set $\Lambda$ of realization of the market, i.e. a set of $\alpha_{i}$, the investor adds a little ``error''$\epsilon_{i}$ on each asset: from the  investor's point of view, the set $\Lambda$ appears to be the set $\Lambda'=\{ \alpha'_{i} = \alpha_{i}+\epsilon_{i} \}_{i}$. The $\epsilon_{i}$ are independent random variables drawn from a Gaussian distribution with zero mean and variance $d$. $d$ is then a measure of the precision of the investor. The smaller is  $d$, the clearer is the investor's perception of the assets. 
One can give a rough estimate of the typical growth rate of an investor who can choose among the $N$ assets to diversify his portfolio.
First of all, the investor will look for assets with positive average return: in his perception, he will then include in his portfolio only the assets with a positive $\alpha'_{i}$. As the $\epsilon_{i}$ and the $\alpha_{i}$ are Gaussian distributed, the $\alpha'_{i}$ will be distributed as a Gaussian variable with mean $m$ and variance $d+D$. Thus, to estimate the number $\overline{n}$ of assets to be included in the portfolio, we will use is expectation value $\overline{n} \simeq NProb(\alpha'>0)$, where $Prob(\alpha'>0)$ is the probability to have a value of $\alpha'$ greater than 0.
Given a set $\alpha_{i}$, we put them in a decreasing order, so that $\alpha_{1}>\alpha_{2}>\ldots>\alpha_{N}$.
One can estimate the value of the $\alpha_{i}$, that we note where now the index $i$ is referred to the order, by means of the condition $P(\alpha>\alpha_{i}) = \frac{i}{N}$, that is
\begin{equation}
\alpha_{i}=m+\sqrt{D}T(\frac{i}{n}).
\end{equation}
where we have noted by  $T(x)$ the function
\begin{equation} \label{igf}
x \rightarrow T(x) \equiv \{ t \in \mbox{I\hspace{-.15em}R} \mid  \int_{-\infty}^{t}\frac{e^{-\frac{s^{2}}{2}}}{\sqrt{2\pi}}ds = x \},
\end{equation}
i.e. the ``inverse function'' of the gaussian cumulative distribution. As in [1], one will allocate a fraction $f_{opt}(\alpha)=\frac{\alpha}{\alpha^{2}+\Delta}$ on an asset with average return $\alpha$ and volatility $\Delta$. The average value (i.e. with respect to the distribution of errors and over all the possible realization of the market) of the typical growth rate of an investor who provides a given effort $E$ is given by eq. \ref{vtip}, who can be written, in case of ``small'' returns, as
\begin{equation}
\overline{\avg{V_{typ}(\Lambda,d(E))}}=\sum_{i=1}^{\overline{n}} [\alpha_{i}\overline{f_{opt}(\alpha_{i}+\epsilon)} - \frac{\overline{f_{opt}^{2}(\alpha_{i}+\epsilon)}}{2}(\alpha_{i}^{2}+\Delta)].
\end{equation}
Again, we expand $f^{opt}(\alpha+\epsilon)$ up to 2nd order in $\epsilon$, and take the average with respect of the distribution of the $\epsilon_{i}$, to obtains the expressions for the moments of $f_{i}$.
If we note $V_{0}=\overline{\avg{V_{typ}(\Lambda,0)}}$, the typical growth rate of an investor provided with a perfect knowledge of the assets, we are then able to write
\begin{equation}
\overline{\avg{V_{typ}(\Lambda,d(E))}}= V_{0} - \frac{d}{2}\sum_{i=1}^{\overline{n}}\frac{(\alpha_{i}^{2}-\Delta)^{2}}{(\alpha_{i}^{2}+\Delta)^{3}}.
\end{equation}
where $d$ is a decreasing function of the effort. So, as expected, $\overline{\avg{V_{typ}(\Lambda,d(E))}}$ is a increasing function of the effort provided to get a better knowledge of the market, as it is shown in Fig. 2 referring to a ten assets market with a average growth rate of $-0.5\% $.

\section{Fixed Investment Fraction}
An investment must be diversified upon different assets, to reduce fluctuation and increase typical growth. But in order to reduce fluctuations, the investor must include less advantageous assets in his portfolio, in a realistic case. This leads to an optimal number of assets included in the portfolio. This mechanism has been described by [1]. If the investment is diversified upon $n$ different assets with the following set of average returns and volatilities $\{\alpha_{i},\Delta\}$ for $i \geq 1$, {\it a priori} the investors should put a fraction $r_{i}=\frac{\alpha}{\alpha^{2}+\Delta}$. We assume that $\Delta + \alpha^{2} \simeq \Delta$ as it is often the case in real finance. Thus, the investor should allocate the capital fraction $r_{i}(\alpha_{i},\Delta) \simeq \frac{\alpha}{\Delta}\theta(\alpha)$ on each asset.
If the total invested fraction is fixed, and equal to $C$, we have to satisfy the condition $\sum_{i=1}^{n}r_{i} = C$. This can be done by introducing a Lagrange multiplier $\lambda(n)$ with the condition $\sum_{i=1}^{n}(r_{i}-\lambda(n))=C$,
and then the optimal fraction to be invested on asset $i$ is $f_{i} = \frac{\alpha_{i}}{\Delta} - \lambda(n)$.
Assets  available on market are not all equal. We suppose to have $N$ assets whose average return and squared volatility are $\{ \alpha_{i}, \Delta \}$. Again, we can reorder the $\alpha_{i}$ to have $\alpha_{1}> \alpha_{2} > \ldots > \alpha_{N}$.
We suppose that the $\alpha_{i}$ are distributed as a Gaussian variable around a mean value $m$ with a square deviation equal to $D$. The mean value acts as an average variation of the market index, and the deviation $D$ is a measure of the market volatility. 
Moreover, investor hasn't got an exact perception of the real value $\alpha_{i}$: in his perception, the i-th assets has an average return $\alpha'_{i}=\alpha_{i}+\epsilon_{i}$, where $\epsilon_{i}$ are drawn from a Gaussian distribution, with mean value $\avg{\epsilon}=0$ and square deviation $\avg{\epsilon^{2}}=d$. The more exact is the perception, the smaller is the $d$. So, in the investor's perception, assets drift values $\alpha'_{i}$ are still distributed as a Gaussian variable, with the same mean value $m$ and variance $D+d$, as it is the variance of the sum of two Gaussian variables. Thus, the investor is led to put on the $i$-th assets a fraction $r_{i}(\alpha'_{i},\Delta)$. 
To have the maximal diversification, the investor has to choose the ``best'' $n$ assets satisfying the condition that the related investment fraction $r_{i}$ is bigger that the Lagrange multiplier $\lambda(n)$ built upon the $n$ assets, as seen in [1]. For a given $n$, we have $n\lambda(n)=\sum_{i=1}^{n}{r_{i}}-C$.
The condition to have $n$ assets included in the portfolio, is that $r_{n}>\lambda(n)$ and $r_{n+1}<\lambda(n+1)$. Then, the optimal value of $n$ is given by the self-consistent equation $n r_{n}=\sum_{i=1}^{n}{r_{i}}-C$.
To evaluate $r_{i}$, we use a rough estimate for $\alpha'_{i}$, that we note $\overline{\alpha'_{i}}$, obtained through the condition $Prob(\alpha'_{i}>{\overline{\alpha'_{i}}})=\frac{i}{N}$. Recalling to the reader the definition (\ref{igf}) of the function $T$, we can write
\begin{equation}
\overline{\alpha'_{i}} = m + \sqrt{D+d}T(\frac{i}{N}).
\end{equation}
By replacing this expression in the self-consistent equation, we obtain
\begin{equation}
\frac{\Delta C}{\sqrt{(D+d)}} + nT(\frac{n}{N}) - \sum^{n}_{i=1}T(\frac{i}{N}) = 0.
\end{equation}
By solving this equation, one obtains an estimate of the number of assets upon which an investor with an ``ignorance'' $d$ will invest.
The dependence on N shows that in a more populated market there will be more assets on which it could be worth to invest, rather than a less populated one, so that the diversification of the portfolio can be made over all ``good'' assets.

\section{Knowledge is a Driving Force: a Single Cell case}
Now we look with more detail the case of an investor forced to put all his capital on two assets ($f_{1}+f_{2}=1$), that we still represent as multiplicative Gaussian random walks
\begin{equation}
p_{i}(t+1)=e^{x_{i}(t)}p_{i}(t),
\end{equation}
for $i=1,2$. We note by $m_{i}$ and $\Delta_{i}$ the mean and the variance of the random walk $x_{i}$, for $i=1,2$. If by $f$ we note the fraction invested on asset ``$1$'', after a 2nd order expansion in $x_{1}$ and $x_{2}$, and assuming that $m_{i}^{2} \ll \Delta_{i}$, we get
\begin{equation}
V_{typ}  \simeq f(m_{1}+\frac{\Delta_{1}}{2})+(1-f)(m_{2}+\frac{\Delta_{2}}{2})-f^{2}\frac{\Delta_{1}}{2} - (1-f)^{2}\frac{\Delta_{2}}{2},
\end{equation}
and it is easy to see that the value of $f$ that maximize the typical growth rate is 
\begin{equation}
f^{opt}(m_{i},\Delta_{i})=\frac{1}{2}+\frac{m_{1}-m_{2}}{\Delta_{1}+\Delta_{2}}.
\end{equation}
We assume that the investor has a wrong perception of the real statistical parameters $m_{1}$ and  $m_{2}$ of the assets. Then, in his perception $m_{1}' = m_{1}+ \epsilon_{1}$ and $m_{2}' = m_{2}+ \epsilon_{2}$ are the mean values of the random walks, then he will allocate a fraction equal to $f^{opt}(m'_{i},\Delta_{i})$.
If $\epsilon_{1}$ and $\epsilon_{2}$ are Gaussian variables with $\avg{\epsilon_{i}}=0$ and $\avg{\epsilon_{i}^{2}}=d$, $V_{typ}$ too becomes a random variable, whose density function is easily calculated by means of standard probability relations for functions of random variable. 
\begin{eqnarray}
P(V) = \frac{1}{Z}\sqrt{\pi S (V_{0}-V)}\exp{-\frac{V_{0}-V}{S}} + A\delta (V-m_{2})+B\delta(V-m_{1}), \nonumber
\end{eqnarray}
where 
\begin{eqnarray}
S & = & \frac{2d}{\Delta_{1}+\Delta_{2}}, \\
A & = & Prob(x < \frac{\frac{\Delta_{1}+\Delta_{2}}{2}+m_{1}-m_{2}}{\sqrt{2d}}), \\
B & = & Prob(x < \frac{\frac{\Delta_{1}+\Delta_{2}}{2}+m_{2}-m_{1}}{\sqrt{2d}}), \\
V_{0} & = & \frac{m_{1}-m_{2}}{2}+\frac{\Delta_{1}+\Delta_{2}}{8}+\frac{(m_{1}-m_{2})^{2}}{2(\Delta_{1}+\Delta_{2})},
\end{eqnarray}
the re-normalization constant $Z$ is such that $\int_{-\infty}^{V_{0}}P(V)dV = 1$, and the probability function appearing in the expression of $A$ and $B$ is referred to a normally distributed random variable.
If $A$ and $B$ are small, the last two terms in the density function can be dropped and we obtain the following derivative, as shown in Fig. 3, where we have fixed the average growth rates equal to $1\% $ (for both assets), and volatilities to $30\% $ (for both assets):
\begin{equation}
\avg{V_{typ}}'(d) = -\frac{1}{\Delta_{1}+\Delta_{2}} < 0
\end{equation}

\section{Investment Network}
We introduce a set of identically distributed multiplicative random walks, to describe price variations:
\begin{equation}
p_{i}(t+1)= e^{x_{i}(t)} p_{i}(t)
\end{equation}
for $i=1,\ldots,N$. We note by $\alpha_{0}$ and $\Delta_{0}$ their average return and volatility, as defined above. 
We note by $ \avg{.}$ the average over the distribution of the $x_{i}$. The web-like model is built in the following way. Starting from $i=1$, each pair of consecutive assets $p_{i}(t)$ and $p_{i+1}(t)$ becomes a two-assets portfolio for an investor, whose capital stock follows a random walk driven by the ``basic'' random walks $p_{i}(t)$. In this way, we obtain a new set of N investor (one for each pair of subsequent assets), whose capitals describes a multiplicative random walk ``driven'' by the basic one. 
Iterating this process of ``two-assets investing'' over this new set of random walk, and assuming periodic boundary conditions (i.e., we add a N+1-th site at each level identical to the first site), we build a network made of subsequent levels of $N$ investors, that we note $W^{(l)}_{i}$ where $l$ stands for the level and $i$ is the site index. The capital of each investors then follows a multiplicative random walk. We introduce a generalized multiplicative step, or ``return'', by the definition
\begin{equation}
R^{(l)}_{i}(t) \equiv \frac{W^{(l)}_{i}(t+1)}{W^{(l)}_{i}(t)}.
\end{equation} 
The set of the $N$ basic random walks {$p_{i}(t)$} is the 0-th level, so that $W^{(0)}_{i}(t) = p_{i}(t)$ and $R^{(0)}_{i}(t) = e^{x_{i}(t)}$.
Even if the model still remains a simple description of economical systems, it takes into account two factors: 1) the need of diversification of a portfolio, in order to absorb the shocks; 2) the impossibility of getting a perfect knowledge about all economic factors, that leads to an upper limit for the diversification itself. If one could get information ``for free'', he could invest on all the favorable assets, (see Fig. 4).
In our model, each agent invests a constant fraction $f$ of his capital in each of the two assets who stand below him. So, investors at the l+1-th level are ruled by the following set of equations of motion:
\begin{equation} \label{eqmot}
W^{(l+1)}_{i}(t+1) = [1-2f+f(R^{(l)}_{i}(t)+R^{(l)}_{i+1}(t))]W^{(l+1)}_{i}(t).
\end{equation}
for $l=0,1,2,\ldots$. From eq. (\ref{eqmot}), it can be easily seen that
\begin{equation} \label{recurs}
R^{(l+1)}_{i}(t) \equiv 1-2f+f(R^{(l)}_{i}(t)+R^{(l)}_{i}(t)).
\end{equation}
and by iteration over all values of $l$, we obtain
\begin{equation}
R^{(l)}_{i}(t)=1-(2f)^{l}+f^{l}\sum_{i=0}^{l} ( \begin{array}{c}l \\ i \end{array}) R^{(0)}_{i}(t). 
\end{equation}
In the following calculation of the average properties values of $R^{(l)}_{i}(t)$, we can drop the index $i$ and the time parameter $t$, since the average properties of all the basic random walks are identical and every step is independent and we only keep the level indexes $l$. After some straightforward calculation, we get exact expressions for the average return $\alpha_{l} \equiv \avg{R^{(l)}}-1$ at the l-th level of the network
\begin{equation}
\alpha_{l} = (2f)^{l}\alpha_{0},
\end{equation}
and for the squared volatility $\Delta_{l} \equiv \avg{(R^{(l)})^{2}}-{\avg{R^{(l)}}}^{2}$
\begin{equation}
\Delta_{l} = f^{2l}\Delta_{0}( \begin{array}{c}2l \\ l \end{array}).
\end{equation}
By these expression, and assuming that $R^{(l)}-1 \ll 1$, we can expand the logarithm of the r.h.s. of eq.(\ref{recurs}) up to 2nd order in $R^{(l)}-1$, and obtain an expression for the typical capital growth rate for an investor belonging to the (l+1)-th level of the network:
\begin{equation}
\avg{\ln{(R^{(l+1)})}}\simeq 2f(2f+1)\avg{R^{(l)}}-2f(1+f)-f^{2}\avg{(R^{(l)})^{2}}-f^{2}\avg{R_{i}^{(l)}R_{i+1}^{(l)}}
\end{equation}
where the last term, that is the correlation between two neighbor sites, is given by
\begin{equation}
\avg{R_{i}^{(l)}R_{i+1}^{(l)}}=[1+(2f)^{2}\alpha_{0}]^{2}+f^{2l}\Delta_{0}\frac{(2l)!}{(l+1)!(l-1)!}.
\end{equation}
One can see in fig. 5, where we have plotted a case in which the basic random walks have an average return of $1 \%$, a volatility of $10 \%$ and agents invest a $20 \%$ fraction of their capital on each assets, that investors belonging to upper levels can take advantage of the lower ones, who dump the fluctuations coming of the basic random walks, but also get a lower gain, since their capital evolution has a weaker dependence on the basic, and hopefully favorable, price variations.

\nonumsection{References}
\noindent

\small\baselineskip=11pt
\halign{\noindent#\hfil&\quad\vtop{\parindent=0pt\hsize=27.5pc
\hangindent .0em\strut#\strut}\cr
[1] &S.Maslov and Y.-C. Zhang, ``Optimal Investment for Risky Assets'', {\em International Journal of Theoretical and Applied Finance}, {\bf 1}, {\em 3} (1998), 377-387.\cr
[2] &E.Elton and M.Gruber, {\em Modern Portfolio Theory and Investment Analysis}, J.Wiley, 1995.\cr
[3] &Y.-C. Zhang, ``Toward a Theory of Marginally Efficient Markets'', {\em Physica A}, 269 (1999), 30-44.\cr
[4] &S.Grossman and J. Stiglitz, ``On the impossibility of Informationally Efficient Markets'', {\em The American Economic Review}, 3 (1980), 393-408.\cr}

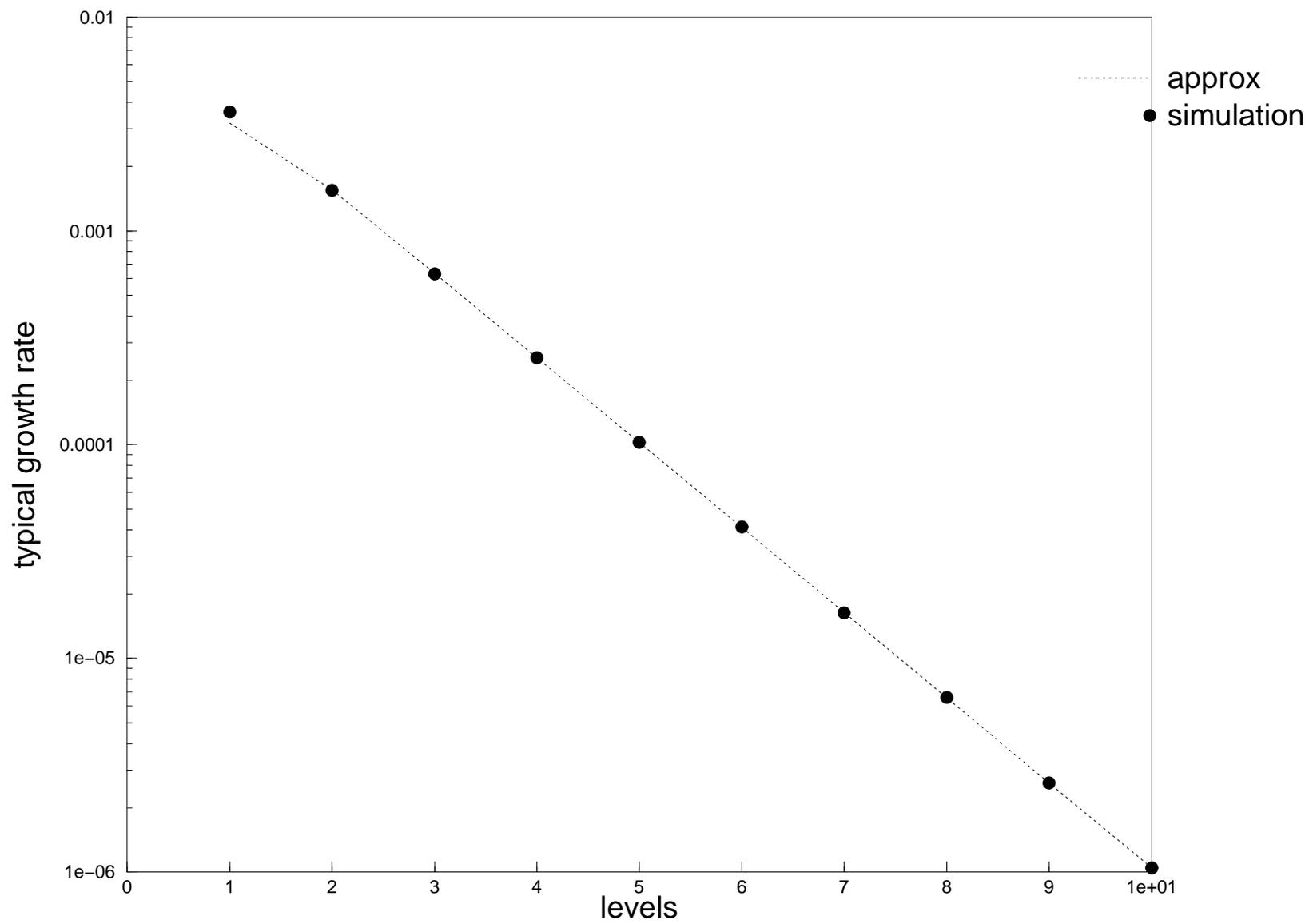
\end{document}